\documentstyle[12pt]{article}
\def\bcc{\begin{center}} \def\ecc{\end{center}}
\def\beq{\begin{equation}} \def\eeq{\end{equation}}

 \def\n{\noindent}

\def\pom{\sl{{\hskip2pt I}{\hskip-2pt P}}\rm \ }
\def\regg{$\hskip4pt I\hskip-4pt R_\rho $}
\def\reg2{\hskip4pt I\hskip-4pt R}

\begin{document}
\vskip-1.5cm
\hskip12cm{\large HZPP-9801}

\hskip12cm{\large Jan 5, 1998}

\vskip1cm
\centerline{\Large\bf THE CONTRIBUTION OF REGGEON}

\vskip0.5cm
\centerline{\Large\bf IN CHARGE EXCHANGE PROCESSES}

\date{}

\vskip0.5cm
\centerline{Zhou Yufeng$^{1}$  \ \ \ \ Peng Hongan$^{2}$ \ \ \ \ 
Liu Lianshou$^{1}$}
\vskip2cm

\bcc {\bf Abstract} \ecc
\bcc
\begin{minipage}{120mm}
\small
We discuss in this paper 
The experimental results on maximum psedo-rapidity
$\eta_{max}$ distribution in the charge exchange process  $e+p\rightarrow
e+n+X$  in ZEUS Collaboration at HERA.
We calculate the contributions of \regg ($\rho$-Reggeon associated 
with $\rho$ meson) from regge phenomenology
and $\pi^{+}$-exchange from pion cloud model.
The results show that
neither the \regg-exchange nor the pion cloud model alone can explain the 
experimental data well, but after considering  both these two processes
 together,
 by using Monte Carlo simulation, a  good agreement between
theoretical results and experimental data is found. This means that in
discussing the  large rapidity gap phenomena in deep inelastic scattering,
 both of the two processes play substantial role.
\end{minipage}
\ecc 
\vskip0.8cm
\n{\bf Keywords: }  Reggeon-exchange,  large rapidity gap, pion-exchange

\vfill
{\small 1 Institute of Particle Physics, Huazhong Normal University, Wuhan
430079 China}

{\small 2 Department of Physics, Peking University, Beijing  100871  China }

\newpage

\section{Introduction}
In recent years, the observation of large rapidity gap (LRG) events from ZUES and H1{\cite{zeus}} 
collaborations in HERA has attracted much interest. This class of events can
not be predicted from standard deep inelastic scattering (DIS) which assumes that
a colored quark was stuck by lepton, so that there is a color field between
the proton remnant and the stuck quark, and  particles will be produced in the 
whole region between them. On the other hand, large rapidity gap means that
there is little energy deposited in the forward region. In terms of rapidity distribution
this means that there is a large gap
between the produced particles and the beam remnant. The ratio of the frequency of
large rapidity gap events to that of normal DIS events is about 10\%.

  	The LRG events suggest that in these processes, a color singlet object is exchanged.
One of the most probable candidate of this object is Pomeron(\pom), ie. the Regge trajectory
with the quantum number of vacuum. In the process of Pomeron exchange the 
proton is diffractivly scattered into the forward region and separated from the other
particles in rapidity. The models based on Pomeron exchange such as the I-S model\cite{IS},
D-L model\cite{DL} have made great success on describing the properties of LRG events.

Since  pomeron is one of the many Reggeon trajectories, one may go
further to ask what about the contribution of other Regge trajectories in
 strong interaction. In $e+p$  forward hard diffractive scattering process,
 Beside the contribution of pomeron, other reggeon exchange such as
$\rho_{0}$-reggeons, and the $\pi^{0}$ meson exchange must be considered,
although their contribution  to the cross section are about one or two order
 of magnitude smaller than that of pomeron exchange.

	But, as we know, there is a significant difference between pomeron and 
other Reggeons. The \pom has the quantum number of vacuum, the other Reggeons have
quantum numbers of the particles which belong to the corresponding Regge
trajectories. So \pom-exchange and Reggeon-exchange may occur in different processes.
If we can isolate some process of Reggeon-exchange from that of \pom ones, we will get
the opportunity of studying the properties of Reggeon-exchange. In \cite{pen} the 
\regg-exchange ( the exchange of $\rho$ -Reggeon associated with $\rho$
meson) was studied. Since the \regg \ has the quantum number of $\rho^+$, the \regg
-exchange process is a charge exchange process. In the case of \pom-exchange, no
charge was exchanged. So in \pom-exchange we will find the outgoing proton in
forward direction. On the contrary, in \regg-exchange the final forward hadron is a
 neutron. Recently a forward neutron calorimeter (FNC) was installed in ZEUS Collab.
in HERA, which can detect the fast neutron in forward region, so the charge exchange
process can be separated from neutral current process.

	In this paper we first present the formalism of \regg-exchange in section 2.
Then, for a comparison the process of $\pi^+$-exchange from pion cloud model 
was also discussed in section 3. In section 4, we use Monte Carlo method to study
 the exclusive properties  of these two process. The $\eta_{max}$ distribution is
 obtained. From the comparison to the experimental data, we find that neither
 the \regg-exchange nor the $\pi^{+}$-exchange alone can describe the data,
 but combining these two processes together, 
the experimental data can be explained very well.

\section{The formalism of \regg-\bf{exchange}}

The \regg-exchange has been studied in $\pi$-$N$ interaction\cite{rev}. In the reaction:
\beq \pi^- + P \rightarrow \pi^0 + N \eeq
only the $\rho^+$-Regge trajectory (with quantum numbers $J^{pc}=1^{--}$)
can be exchanged. In Regge theory the $s$-channel helicity amplitudes has the
form\cite{collins}:
\beq
A_{H_s}(s,t)\stackrel{s \gg m^2_N}{\sim}(\frac{-t}{2 m^2_N})^{1\over2 m}
(\frac{1-e^{-i\pi\alpha_\rho(t)}}{\sin \alpha_\rho(t)})
(\frac{s}{2m^2_N})^{\alpha_\rho(t)}\gamma_\pi(t)\gamma_{\mu_2\mu_4}(t) ,
\eeq
where $A_{H_s}$ is the $s$-channel helicity amplitudes, $s$ is the energy squared in
cms frame, $m_N$ is the mass of nucleon, $t$ is the four momentum transfer squared,
$\alpha_\rho(t)$ is the Regge trajectory of $\rho^+$. 
$\gamma_\pi,\gamma_{\mu_2 \mu_4}(t)$ 
are vertex functions of \regg$\pi\pi$ and \regg$NN$ respectively.
$m$ is defined as:
\beq m\equiv \mid \mu_2 -\mu_4 \mid . \eeq
$\mu_2,\mu_4$ are helicities of the initial proton and final neutron, see Fig.1. Since in 
diffractive interaction $t$ is very small, we set $m$=0 \cite{pen}.
The differential and total cross section can be written as:
\beq
\frac{d\sigma}{dt}(s,t)=\frac{1}{64\pi s k^2}\frac{1}{2}\sum |A_{H_s}(s,t)|^2 ,
\eeq 
\vskip10pt
\beq
\sigma_{tot}=2\beta_\rho^{2}(\frac{s}{2 m_N^{2}})^{\alpha(0)-1} \\
\eeq
respectively,
where $k\equiv p^\pi_{c.m}$ is the momentum of pion in $\pi$-$N$ cms frame,
$\beta_\rho$ is the coupling function between
\regg \ and light quarks (u,d quarks).
From the $\pi$-N phenomenology the $\beta_\rho$ can be estimated as\cite{pen}:
\beq \beta_\rho(t)|_{t=0}=2.07 \ {\rm GeV}^{-1} .   \eeq
The $\alpha_\rho(t)$ can be written approximately as:
\beq \alpha_\rho(t)=\alpha_\rho(0)+\alpha_\rho^{'} t , \eeq
where $\alpha_\rho(0)\approx0.5, \alpha_\rho^{'}\approx0.9$  GeV$^{-2} $.
Thus, the effective propagator of \regg \ is :
\beq
D_\rho(t)=2\beta_\rho^{2}F_N(t)(\frac{s}{2 m_N^2})^{\frac{1}{2}(\alpha_\rho(t)-1)}
(\frac{1-e^{-i\pi\alpha_\rho(t)}}{\sin\pi\alpha_\rho(t)}) ,
\eeq
where $F_N(t)$ is the form factor for a quark in nucleon. It can be parameterized as follow:
\beq F_N(t)=\frac{4m_N^2-2.8t}{4m_N^2-t}(\frac{1}{1-\frac{t}{0.7}})^2 . \eeq
Like the case of pomeron, it is also possible for the Reggeons to have partonic structure, i.e. 
 it would appear in hard subprocess. In \cite{pen}, the process sketched in
Fig.2a was evaluated
by using effective propagator and the coupling constant of  \regg. The problem was handled
very similar to the D-L model for \pom, where the coupling between \pom and the 
quarks is point-like.
In the process of charge-exchange photo production quark-pair process : 
\beq \gamma^{*}+P\rightarrow n+q_u+\stackrel{-}{q}_d , \eeq
the invariant amplitude is expressed  as :

\beq
M=D_\rho(t)[Q_u \stackrel{-}{u}_u(k)\gamma_\mu \frac{i(\not q^{'}-\not k^{'})}{(q^{'}-k^{'})^{2}}
\gamma_\rho v_d(k^{'})+Q_d \stackrel{-}{u}_u(k)\gamma_\rho \frac{i(\not q-\not k^{'})}{(q-k^{'})^{2}}
\gamma_\mu v_d(k^{'})]\stackrel{-}{u}_d(p^{'})\gamma_\rho u_u(p) ,
\eeq
where $Q_u=\frac{2}{3}e,Q_d=\frac{-1}{3}e $ is the charge of $u,d$ quark respectively.
The corresponding cross section $\sigma_{\gamma^{*}P}^{T}$ is :
\beq
\sigma_{\gamma^{*}P}^{T}(s,Q^{2})=\frac{1}{4 q_{s12}\sqrt{s}}
\int \overline{\sum}|M|^{2}(2 \pi)^{4}\delta^{4}(p+q-p^{'}-k-k^{'})
\frac{d^{3}p^{'}}{(2\pi)^{3}2E_{p^{'}}}
\frac{d^{3}k}{(2\pi)^{3}2E_{k}}
\frac{d^{3}k^{'}}{(2\pi)^{3}2E_{k^{'}}} .
\eeq
Thus, the contribution to the structure function $F_2(x,Q^{2})$ can be evaluated
under the equivalent photon approximation:
\beq
F_2(x,Q^{2})=\frac{Q^{2}}{4\pi^2 \alpha_{em}} \sigma_{\gamma^{*}P}^{T}(s,Q^{2}) .
\eeq

\section{The $\pi^{+}$-exchange in pion cloud model}
   Recently the concept of pion cloud in nucleon becomes very
successful in understanding the Grotfried sum rule violation observed by the
New Muon Collaboration and the Drell-Yan asymmetry measured in NA51 at CERN
\cite{NMC}.

   In the charge exchange process, it is well known that the $\pi^{+}$-exchange
process of pion cloud model\cite{cloud}
gives the largest contribution. The formalism of pion-exchange was proposed many
years ago\cite{salivn}. In recent years, since FNC has been installed on ZEUS Collab., 
one can study this process experimentally and get a very good chance to study 
the structure of pion. Some methods have
been proposed to measure the pion structure function on HERA\cite{pl94}. 

The process
of $\pi^{+}$-exchange is sketched in Fig.3a and Fig.3b. The cross section is factorized 
into \cite{zphy}:
\beq
\frac{dF_2^{p}}{dt dz}=f_{\pi N}(z,t)F_2^{\pi}(x_\pi,Q^{2}) ,
\eeq
where $f_{\pi N}(z,t)$ is the flux factor, $F_2^{\pi}$ is the structure 
function of pion, $z$ is the ratio of the neutron  to proton light-cone momenta,
$x_\pi $ is the Bjorken variable for the pion, $x_\pi=x/(1-z)$. The flux
factor for Fig.3a can be expressed as \cite{zphy} :
\beq
f_{\pi N}(z,t)=\frac{2g_\pi^{2}}{16\pi^{2}} \frac{|t|}{(m_\pi^{2}-t)^{2}}
G_1^{2}(t)(1-z) .
\eeq
In\cite{zphy}, the factor$(1-z)$ is changed into $(1-z)^{1+2\alpha_\pi^{'}|t|}$.
where $\alpha_\pi^{'}$ is the slop of the $\pi$-Regge trajectory, $\alpha_\pi
^{'}\approx 1$.
 This approach is wildly used to 'reggeize' the $\pi$-exchange process when
$(1-z)$ is very small or $|t|$ is very large\cite{Ing}\cite{zphy}. In 
this kinematical region the off-shell behavior of pion should be considered
\cite{pumplin}. But since the $\pi$-exchange is in the kinematical
region $z \approx 0.7-0.9$ and $|t| \leq 0.2 \approx 0.4$ GeV$^2$\cite{zphy}, 
 the influence of 'reggeize' is insignificant in this process.
 So we choose $(1-z)$ instead of $(1-z)^{1+2\alpha_\pi^{'}|t|}$,
in consistence with the on shell-pion structure function, going to be used
in the following discussion,  
Because the difference is important only in very limited region of
phase space, both approaches lead to almost the same result\cite{Ing}\cite{zphy}.

The value of $\pi NN$ coupling constant $g_\pi$ is fixed at 
$g_\pi^{2}/ 4\pi=13.75$. 
$m_\pi$ is the mass of pion, $G_1(t)$ is the form factor, which
has the form $G_1(t)=exp[R_1^{2}(t-m_\pi^{2})]$, where $R_1^{2}=0.3$ GeV$^{-2} $.
The flux factor of Fig.3b has different form\cite{zphy}. From\cite{zphy}, we know that
the contribution of Fig.3b is less than that of Fig.3a for about  one order of magnitude. 
It will not be so important in the following di
scussion, so we neglect it.
The structure function $F_2^{\pi}(x_\pi,Q^{2})$ can be determined through a fit to
experimental data as\cite{zphy}\cite{F2}:
\beq
F_2^{\pi}(x,Q^{2})=\frac{2}{3}f(Q^{2})[a exp(2\sqrt{L})/L+b\sqrt{x}].
\eeq
Where $L=(4\pi/\beta_{0})ln(c/\alpha_{s})ln(d/x),f(Q^{2})=Q^{2}/(e+Q^{2}),
\alpha_{s}=4\pi/(\beta_{0}ln(Q^{2}/\Lambda_{QCD}^{2}))$.
 The resulting parameters are: $\Lambda_{QCD}=0.2$ GeV, $\beta_{0}=9, a=0.036, b=0.4,
 c=0.59, d=0.31, e=0.12$ GeV$^{2}$.

It is not difficult to understand why the cross section of pion-exchange is
much larger than that of \regg. Firstly, the coupling constant $g_{\pi}$ is
much larger than $g_{\rho}$, their ratio $g_{\pi}^{2}/g_{\rho}^{2}=32.3$. 
Secondly, from (14) and (16) we can see that the cross
section of pion-exchange increases with the increasing of
$cms$ energy $s$ ($s\approx Q^{2}/x$), 
because $F_{2}^{\pi}$ increases when $x$ becomes smaller. On the
contrary, from Regge theory the cross section of \regg-exchange decreases quickly
 when $s$ is increasing, i.e. $\sigma \sim s^{\alpha_{\rho}(0)-1}=s^{-0.5}$.

For the same reason, we can also see the large difference between pion
cloud model and the $\pi$-Reggeon exchange, since the intercept of $\pi$-
Reggeon is less than zero ($\alpha_{\pi}(0)\approx -m_{\pi}^{2}$), the cross
section of $\pi$-Reggeon exchange decreases faster than that of \regg-Reggeon
exchange, so that its contribution will be very small at high energy.

\section{Hadronization}
 
In the above sections, we have discussed the inclusive properties of \regg-exchange
and pion-exchange. It is more attractive to ask what can be predicted from these theories 
on the exclusive properties of final hadrons. Since both \regg \ and pion are color
singlet object, one may expect the large rapidity  gap will appear in $\eta_{max}$
distribution, here $\eta_{max}$ is the maximum pseudo-rapidity of a event. From the
theoretical point of view \cite{zphy}, the rapidity gap between the observed neutron
and the hadronic debris can be estimated as: $\bigtriangleup\eta\sim$ ln$(1/(1-z))$. 
Since in $\pi$-exchange, $z$ has the mean value of $z\approx 0.7\sim 0.9$, so 
$\bigtriangleup\eta \sim 1$. It is very small.
We will discuss this problem by using Monte Carlo method. We use Lund string
fragmentation model to describe the fragmentation and hadronization process,
The model has been implemented into the program  JETSETS7.4 and PYTHIA5.7\cite{sjo}.
We use these programs to simulate the process~ :
\vskip-0.2truecm
\beq q_u+q_{\bar{d}} \rightarrow hadrons, \eeq

which is sketched in Fig.2b, and :
\vskip-0.2truecm
\beq \pi^{+} + e \ \rightarrow hadrons  \eeq
\vskip-0.2truecm
\noindent or :
\beq \pi^{+} + \gamma^{*} \ \rightarrow hadrons \\\ (\ for \ fixed \ Q^{2} \ ). \eeq
It is wildly used to use the on shell pions instead of virtual pions in 
Monte Carlo simulation.\cite{Ing}

\section{Results and discussions}

  The Monte Carlo simulation was made corresponding to the HERA condition, i.e.
 $E_p=820$ GeV, $E_e$=26.7 GeV. For the study of the behavior under different
$x$ and $Q^{2}$, the $\gamma^{*}$-$\pi$ interaction was simulated instead of $e$-$\pi$
interaction from (14). We can integrate the integrand over $dt$ and $dz$, then the 
contribution of $\pi$-exchange to the proton structure function $F_{2(\pi)} ^{p}$
can be evaluated. It is very easy from(13) to get the same contribution of \regg
 \ exchange, and their ratio $F_{2 (\rho-exchange)}^{p}/F_{2 (\pi-exchange) }^{p}=0.28$ when 
$x=10^{-3}, Q^{2}=10$ GeV$^{2}$. 
When $x$ goes smaller, this value becomes a little bit smaller too. Note
that this ratio is also the ratio of the differential cross section. From (13) we get:
\beq
\frac{ F_{2 (\rho-exchange)}^{p}}{F_{2 (\pi-exchange) }^{p}}=
\frac{d\sigma_{\gamma^{*}\rho}/dxdQ^{2}}{d\sigma_{\gamma^{*}\pi}/dxdQ^{2}} .
\eeq
By using Monte Carlo method, we can get many properties of these two processes. But
our special interest is in the $\eta_{max}$ distribution. By varying $x$ and $Q^{2}$, 
we get different $\eta_{max}$ distributions of \regg-exchange. It is shown in
Fig.4
that the distribution moves down to the low $\eta_{max}$ region when the Bjorken  
variable $x$ becomes small. But in the case of $\pi$-exchange there is
very little influence (in the region of $10^{-4}<x<10^{-2},10$ GeV$^{2}<Q^{2}<100$ GeV$^{2}$).

  For the comparison to the experimental data measured on FNC \cite{FNC}, the 
simulation was made under the cut: $E_n>$400 GeV, where $E_e$ is the energy of the outgoing
 neutron. In\cite{FNC} the $\eta_{max}$ distribution of the neutron tagged 
events with $E_n>400$ GeV is given. One can find that the $\eta_{max}$ distribution
is very similar to that of normal DIS added by \pom-exchange. But as we know, in 
neutron tagged events the contribution of the \pom vanishes (\pom has the
quantum number of the vacuum). In the experiment on FNC about 3\% events have large
rapidity gap. It seems a little smaller than that of \pom-exchange (\pom-exchange
 gives a contribution  of about 5-7\% to the normal DIS). But it still need a 
new idea to understand it. In order to understand the source of LRG in neutron
tagged events, it is very natural to suggest that \regg-exchange plays an important 
role. From the above discussion, the contribution of the \regg-exchange to the $\pi$
-exchange process has the magnitude of about 3\%. From Fig.4 it is also very clear that
\regg-exchange events have  large rapidity gap. The $\eta_{max}$ distribution of 
\regg-exchange and $\pi$-exchange (the $x$ and $Q^{2}$ are $x=0.0005,Q^{2}=50$ GeV$^{2}$)
 are superimposed in Fig.5 where the experimental
 data of FNC is also given. The region of $\eta_{max}$ shown in \cite{FNC} 
is: $-2.5<\eta_{max} <7.5$, but the region covered by the calorimeters in ZEUS
and can be really measured is $-3.8<\eta_{max}<4.3$.
The values of $\eta_{max}>4.3$ are 'artifact' of the clustering algorithm,
they are strongly affected by limited acceptance towards the forward beam hole.
 In order to avoid the complexity of 'artifact',
 the region of $\eta_{max}$ in Fig.5 is restricted as:
$-2.5<\eta_{max}<4.3$. From Fig.5 we can see that the $\pi$-exchange alone failed
to describe the data. But together with the contribution of the
\regg-exchange, the situation becomes very different. 
 The sum of the contributions from the two processes is shown in Fig.6. It 
can be seen from the figure that a very good agreement between theoretical
 prediction and experimental data is found.\\

In this paper we presented the contribution of \regg-exchange
in charge exchange process. The ratio of the cross section of
\regg-exchange process
to that of $\pi$-exchange is obtained.
We got the $\eta_{max}$ distribution of this process by Monte Carlo simulation 
and compared
it with that of $\pi$-exchange from pion cloud model. We found that neither
one of the two process alone can describe the experimental data from FNC, but 
their sum coincides with the data very well. 
This shows that in studying the 
large rapidity gap phenomena of DIS the contribution from
Reggeon exchange should be considerd\cite{pen}. 
It is especially important in charge exchange processes. 

\noindent
{\bf Acknowledgment: } \\
This work is supported in part by National Nature Science Foundation of China
and the Doctoral Program Foundation of Institution of Higher Education of China.
One of the author (Y.F.Zhou) is grateful to Dr. Cong-Feng Qiao for his kindly
help on the calculation of \regg-exchange, and for his helpful comments.

\newpage

\noindent{\Large\bf Figure Captions}
\vskip 0.3cm
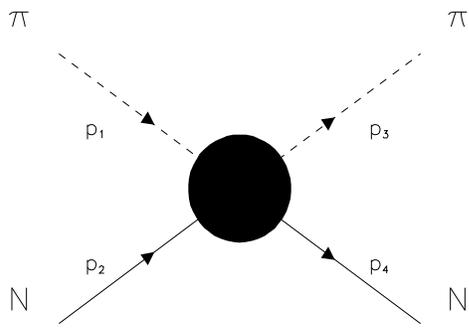
\noindent Fig.1: The scattering processes of $\pi+N \rightarrow \pi+N$.\\
\noindent Fig.2a: The Feynman diagrams for \regg-exchange.\\
\noindent Fig.2b: The process of u, d quarks fragments into hadrons.\\
\noindent Fig.3: Direct (Fig.a) and indirect (Fig.b)  forward neutron 
production by $\pi$-exchange in the pion cloud model.
The dashed line parallel to the solid line at the bottom sketches the
pion cloud.\\
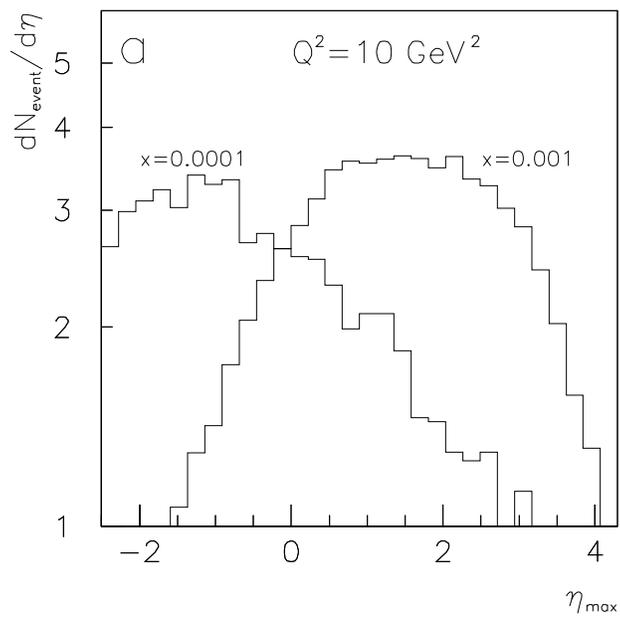
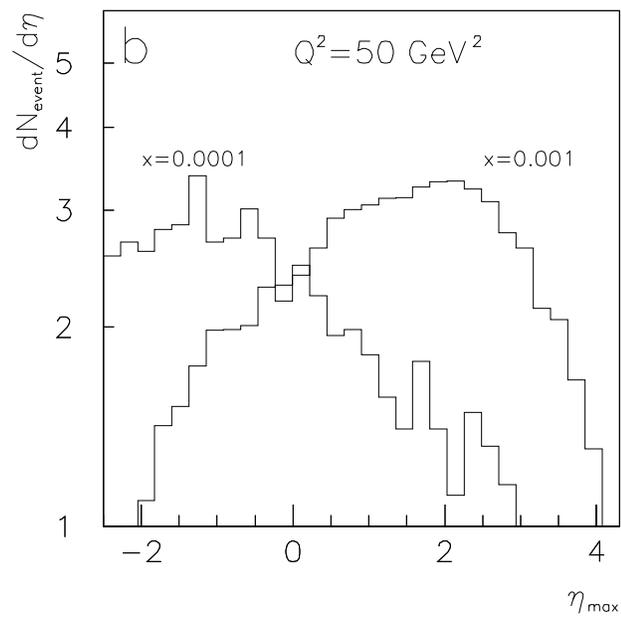
\noindent Fig.4: The $\eta_{max}$ distribution of \regg-exchange in different
$x$ and $Q^{2}$.\\
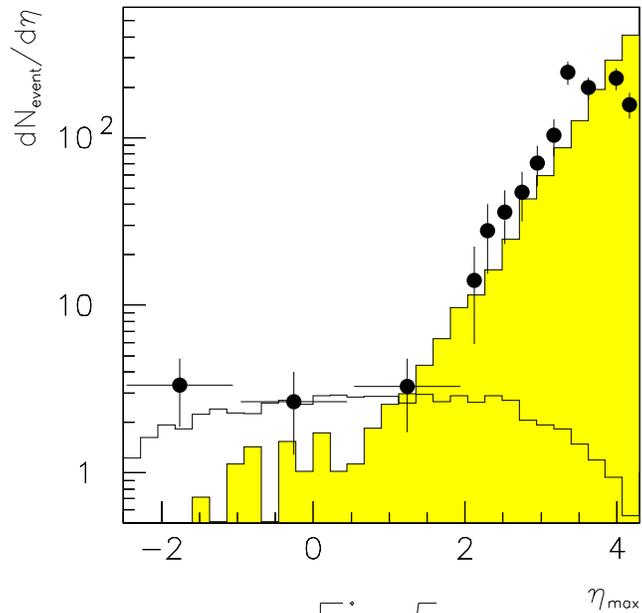
\noindent Fig.5: The $\eta_{max}$ distribution: the dots with err bars are the
experimental data from FNC, the hatched area is the $\eta_{max}$ distribution
of $\pi^{+}$-exchange, normalized to the total dimension of the data with $\eta
_{max}>1.0$, the solid line is the $\eta_{max}$ distribution of \regg-exchange.\\
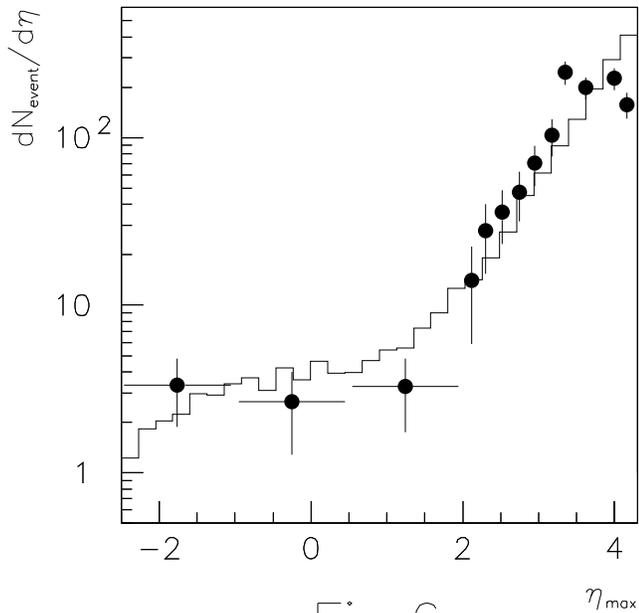
\noindent Fig.6: The $\eta_{max}$ distribution: the dots with err bars are the
data,the solid line is the sum of the contribution of $\pi$-exchange and \regg-
exchange to the $\eta_{max}$ distribution.\\

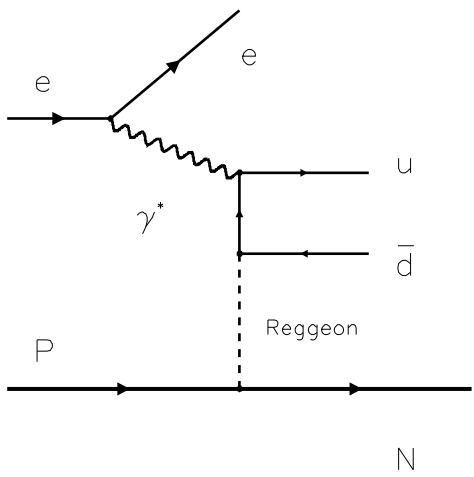
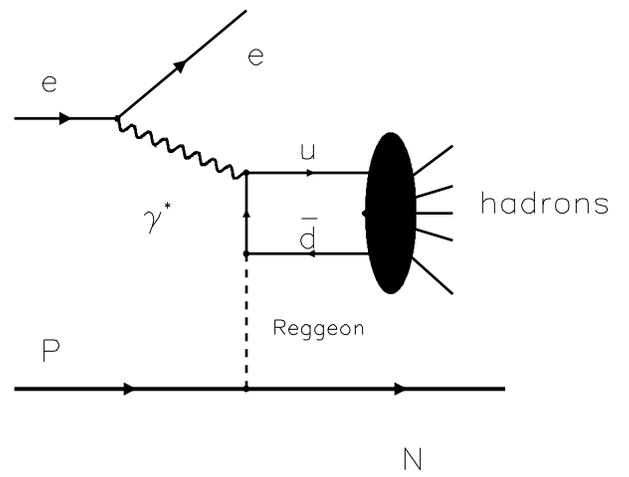
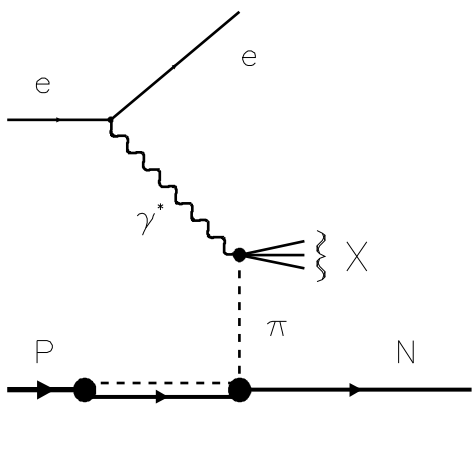
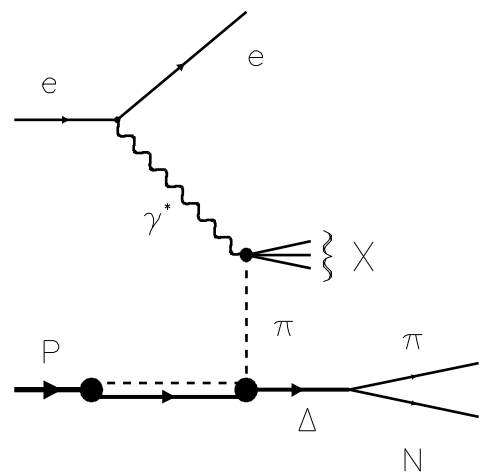
\end{document}